# GAS DAMPING COEFFICIENT RESEARCH FOR THE MEMS COMB LINEAR VIBRATION GYROSCOPE

*Guo Qiufen, Ge Yuansheng, Sun Feng, Liu Fuqiang*

(College of Automation, Harbin Engineering University, Heilongjiang Harbin 150001)

## ABSTRACT

Silicon-MEMS gyroscope is an important part of MEMS ( Micro Electrical Mechanical System). There are some disturb ignored in traditional gyroscope that must be evaluated newly because of its smaller size (reach the level of micron). In these disturb, the air pressure largely influences the performance of the MEMS gyroscope. Different air pressure causes different gas damping coefficient for the MEMS comb linear vibration gyroscope and different gas damping coefficient influences the quality factor of the gyroscope directive. The quality factor influences the dynamic working bandwidth of the MEMS comb linear vibration gyroscope, so it is influences the output characteristic of the MEMS comb linear vibration gyroscope. The paper shows the relationship between the air pressure and the output amplified and phase of the detecting axis through analyzing the air pressure influence on the MEMS comb linear vibration gyroscope. It discusses the influence on the distributive frequency and quality factor of the MEMS comb linear vibration gyroscope for different air pressure.

## 1. BASIC STRUCTURE OF THE MEMS COMB LINEAR VIBRATION GYROSCOPE

Fig. 1 is the sketch of the MEMS comb linear vibration gyroscope. It mainly consists of two side drivers, one mass, damper including springs that jointed the moving mass and static spring, and datum seat. There are some fixed combs in drivers and some moving combs in masses. The two side drivers are fixed on datum seat. The moving mass is jointed with datum seat by flexible springs. The mass are only moved linearly in the direction of x-axis and y-axis due to the action of flexible supports and springs.

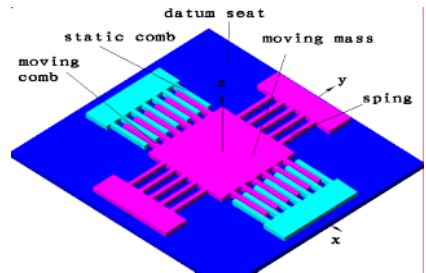

Fig. 1 Sketch of the MEMS comb linear vibration gyroscope

It has periodical electrostatic force between the fixed combs and the moving combs in the direction of x-axis when the moving combs on detecting mass is jointed the ground and a alternating current with direct current excursion is added on the fixed combs on two side drivers. So the detecting mass is vibrated in line periodically between the two side drivers. Here, if the body is tuned round y-axis, the detecting mass is vibrated in line in the direction plumbed the datum seat (i.e. the direction of z-axis).

Because it has sensitivity electrode under the detecting mass, the sensitivity capacitance is changed due to the linear vibration of the detecting mass in the direction of z-axis. The changing quantity of the capacitance is in proportion to the input angle rate (i.e. the angle rate of the body is tuned round y-axis). Then we can get the input angle rate when we detect the changing quantity of the sensitivity capacitance.

## 2. MOVING MODULE OF THE MEMS COMB LINEAR VIBRATION GYROSCOPE

We assume that the angle rate of the body tuned round y-axis is $\omega_{IGy}$. The linear vibrated speed of the moving mass in the direction of x-axis is $v(t)$. Then the Coriolis force acceleration generated in the direction of z-axis is

$$a_z = 2\omega_{IGy} v(t) \quad (2.1)$$

The corresponding rate is

$$v_z = \int 2\omega_{IGy} v(t) dt = 2(\omega_{IG})_y x(t) \quad (2.2)$$

Then the movement equation of the moving mass in the direction of z-axis is

$$z(t) = \int 2\omega_{IGy} x(t)dt = 2\omega_{IGy} \int x(t)dt \quad (2.3)$$

The electrostatic force added on the moving mass in the direction of x-axis is

$$F_l = \frac{(2n-1)\varepsilon_r \varepsilon_0 wV^2}{2d} \quad (2.4)$$

where: n is the number of the moving combs; V is the adding voltage. Because the adding voltage is the alternating current with direct current excursion, the expression of the power voltage is $V = V_0 \pm V_a \sin(\omega_c t)$; w is the overlap width of the comb; d is the clearance between the static and the moving combs; $\varepsilon_r$ is the relative constant of insulated medium; $\varepsilon_0$ is the vacuum relative constant ($\varepsilon_0 = 8.85 pF/m$).

So the movement equation of the moving mass in the direction of x-axis is

$$m\frac{d^2x(t)}{dt^2} + c\frac{dx(t)}{dt} + kx(t) = F\sin\omega_c t \quad (2.5)$$

where: c is the damping coefficient, k is the elastic coefficient, $F = \frac{2(2n-1)\varepsilon_r \varepsilon_0 wV_0 V_a}{d}$.

## 3. GAS DAMPING INFLUENCE FOR THE MEMS COMB LINEAR VIBRATION GYROSCOPE

When the inner configuration of the MEMS comb linear vibration gyroscope is in vacuum, the damping coefficient is zero. Equation (2.5) is changed to

$$m\frac{d^2x(t)}{dt^2} + kx(t) = F\sin\omega_c t \quad (3.1)$$

So the output of the driving mode is

$$x(t) = \frac{F\sin\omega_c t}{m(\omega_c^2 - \omega_n^2)} \quad (3.2)$$

Where, $\omega_n$ is the resonance frequency, it is determined by the mass and the flexibility coefficient and expressed as $\omega_n = \sqrt{k/m}$.

The output of the detecting mode is

$$z(t) = \frac{F(\omega_{IG})_y \sin\omega_c t}{\omega_c(\omega_c^2 - \omega_n^2)} \quad (3.3)$$

When the air is included in the inner of the MEMS comb linear vibration gyroscope, according to the analyses method of amplitude modulation of carrier wave, the output of the driving mode is

$$x(t) = \frac{F\sin(\omega_c t - \Phi)}{\sqrt{(\omega_c^2 - \omega_n^2)^2 + (2\xi\omega_n\omega_c)^2}} \quad (3.4)$$

where: $\xi = \frac{c}{2\sqrt{mk}}$.

Integrated (1.3) and (1.6), we can achieve the output of the detecting mode is

$$z(t) = \frac{-F\omega_{IGy}\cos(\omega_c t - \Phi)}{\omega_c\sqrt{(\omega_c^2 - \omega_n^2)^2 + (2\xi\omega_n\omega_c)^2}} \quad (3.5)$$

$$\tan\Phi = \frac{\omega_n\omega_c}{Q(\omega_n^2 - \omega_c^2)} \quad (3.6)$$

where: $Q = \frac{1}{2\xi}$.

We can see from (3.1)-(3.5) that the pressure is important to MEMS linear vibration gyroscope. The change of the pressure will bring the change of the damping coefficient. The gas created by other matters volatilization and part's surface overflow is increased when the temperature increase. So the density of the gas and the pressure are changed. The relationship of the mucosity coefficient of the gas medium and the temperature is as follows (the other influence of the temperature for the MEMS comb linear vibration gyroscope is not considered here):

$$\mu = \mu_0 (T/T_0)^n \quad (3.7)$$

Where: $\mu_0$ is the gas mucosity coefficient at temperature $T_0$, $\mu$ is the gas mucosity coefficient at temperature T, n is the grade, different gas has different grade.

So, the expression of $\xi$ in (3.4) is as follows:

$$\xi = \frac{c_0 + \mu_0(T/T_0)^n}{2\sqrt{mk}} \quad (3.8)$$

where, $c_0$ is others damping coefficient except gas.

## 4. SIMULATION

Enactment data are as follows:
The length of the comb is $200\mu m$, the width of the comb is $20\mu m$, the space between the moving and the static comb at axis y is $3\mu m$. The resonance frequency is 500Hz, so $\omega_c = 2\pi \times 500 Hz$.
Simulation figures are as follows:

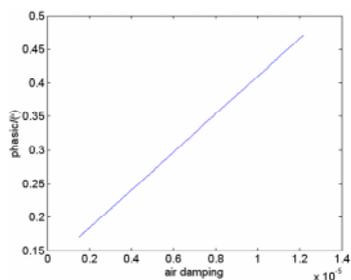

Fig.2 Relationship of the phase and the air damping

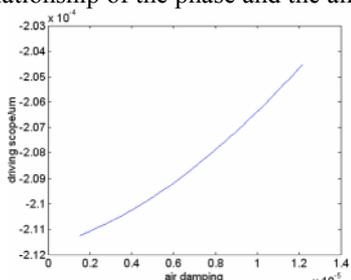

Fig.3 Relationship of the driving displacement and the air damping

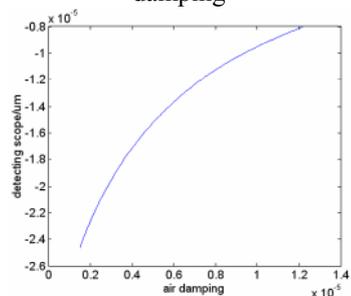

Fig.4 Relationship of the detect scope and the air damping

Figure 2 is the relationship of the phase and the gas damping. We can see from the figure that the relationship of the phase and the gas damping is linearity at most. Its changing scope is 0.17°-0.47°. This changing is arisen by the changing of the gas damping because of the increased temperature only. So influence of the gas damping on the phase is important.

Figure 3 is showed the relationship of the scope of the driving axis and the gas damping. We can see from the picture that the relationship of the driving scope and the gas damping is nonlinear. The driving scope is devalued along with the gas damping increased.

Figure 4 is the relationship of the detecting scope and the gas damping. The picture shows that the relationship of the detecting scope and the gas damping is nonlinear. The detecting scope is devalued along with the gas damping increased.

## 5. CONCLUTION

The changing of the gas medium is one of the most important influence factors for the MEMS comb linear vibration gyroscope. At the same time, it is one of the most neglected factors. We can see from the above analyze that the influence of the gas damping for the MEMS comb linear vibration gyroscope is very great. The size of the damping coefficient decides the characteristic of the frequency response.

## 6. REFERENCES

*Example of a reference of an article from a journal*
[1] Cai Tijing, " Dynamics of Vibratory Wheel Silicon Micromechanical Gyroscope, " *Journal of Southeast University*, Vol.19, No.4, 1999.

*Example of a reference of an article from a journal*
[2] Chen Hong and Bao Minhang, " Air Damping of Micromechanical Silicon Accelerometers, " *Chinese Journal of Semiconductors*, Vol.16, No.12, 1995, pp. 921-927.

*Example of a reference of an article from a journal*
[3] Kan Jouwu, Pen Taijiang, Doung Jingshi, Yang Zhigang and Wu Boda, " Liquid Added Damping and Its Influence on the Output Performance of Micropumps," *Journal of Xi'an Jiaotong University*, Vol.39, No.5, 2005, pp. 548-550.

*Example of a reference of an article from a journal*
[4] Qi Shemiao, Geng Haipeng and Yu Lie, " Effects of the Perturbation Frequency on the Dynamic Stiffness and Dynamic Damping Coefficients of Aerodynamic Bearings," *Journal of Xi'an Jiaotong University*, Vol.40, No.3, 2006, pp. 270-274.

*Example of a reference of an article from a journal*
[5] Ding Siyuan, " THE EFFECTS OF VISCOUS FLUIDS ON ORIGINAL FREQUENCY AND DAMPING OF STRUCTURE," *Journal of Zhengzhou Institute if Light Industry*, Vol.9, No.4, 1994, pp. 50-54.

*Example of a reference of an article from a journal*
[6] Ling Linben and Chen Renkang, " Improve the Tuned Flexure Gyro's Property by Adjusting its Inside Gas Dynamic Pressure Torque," *Journal of Chinese Inertial Technology*, No.2, 1998.

*Example of a reference of an article from a Proceeding*
[7] Timo Veijola, Tapani Ryhämen, Heikki Kuisma and Juha Lahdenperä, " CIRCUIT SIMULATION MODEL OF GAS DAMPING IN MICROSTRUCTURE WITH NONTRIVIAL GEOMETRIES," *The 8[th] International Conference on Solid-State Sensors and Actuators*, and Eurosensors , Stookholm, Sweden, 1995, pp. 36-39.

*Example of a reference of an article from a journal*

[8] Timo Veijola and Marek Turowski, " Compact Damping Models for Laterally Moving Microstructures with Gas-Rarefaction Effects, " *Journal of Microelectromechanical System*, Vol.10, No.2, 2001, pp. 263-273.

*Example of a reference of an article from a Proceeding*

[9] Timo Veijola, Heikki Kuisma and Juha Labdenperä, " MODEL FOR GAS FILM DAMPING IN A SILICON ACCELEROMETER," *1997 International Conference on Solid-State Sensors and Actuators*, Chicago, 1997, pp. 1097-1100.

*Example of a reference of an article from a Proceeding*

[10] C. Bourgeois, F. Porret and Hoogerwerf, " Analytical Modeling of Squeeze-Film Damping in Accelerometer, " *1997 International Conference on Solid-State Sensors and Actuators*, Chicago, 1997, pp. 1117-1120.

*Example of a reference of an article from a journal*

[11] Fan Maojun, Wang Jinsong, Chen Lijie, Li Huimin, Wu Yalin and Wang Ping, " Influence of Damping on Amplitude Frequency Characteristics of Acceleration Transducer," *Journal of Transducer Technology*, Vol.18, No.1, 1999.